# Drop size characteristics of sprays emanating from circular and non-circular orifices in the atomization regime

K. R. Rajesh,[1,2,*] V. Kulkarni,[3] S. K. Vankeswaram,[1] R. Sakthikumar,[1,#] and S. Deivandren,[1]

[1]Department of Aerospace Engineering, Indian Institute of Science, Bengaluru 560012, India

[2]Liquid Propulsion Systems Centre, ISRO, Bengaluru 560008, India

[3]Harvard University, Cambridge 02138, USA

**email ID of all authors**

V. Kulkarni: varun14kul@gmail.com

S.K. Vankeswaram: saikrishna@iisc.ac.in

R. Sakthikumar: sakthi@orbitaid.com

S. Deivandren: dskumar@iisc.ac.in

_______________________________________________________

[*]Corresponding author, K.R. Rajesh's e-*mail address*: rajeshkr@iisc.ac.in

[#]Currently at Orbit Aid Aerospace Pvt. Ltd., Bengaluru 560067, India

## Abstract

Traditionally, circular orifices have been used for generating aerosols, however in recent times non-circular orifices are being considered due to their superior atomization and mixing features. In this work, we experimentally investigate spray morphology and drop size characteristics of kerosene (Jet A-1) jets issuing from three non-circular orifices geometries (elliptic, triangular and, square) and one circular orifice with the same exit cross sectional area. Our results show an unexpected and yet unreported coarsening of atomization for non-circular orifice jets quantified by an increase in the Sauter Mean Diameter (SMD) at all tested exit velocities represented by the liquid Weber number, $We_l$. We attribute this to two distinct spray morphologies: filament and core breakup which generate large size liquid structures identified as filaments and ligaments, noticeable in non-circular orifices jets compared to circular orifice jets. On exploring this further by undertaking a detailed examination of the drop size volume probability distribution at lower $We_l$ corresponding to the location marking the end of primary breakup we see a bimodality due to a dominant distribution of fragments of small and larger sizes owing to the spray morphology. At higher $We_l$ and larger distances from the injector exit we observe the bimodality converts to a unimodal distribution for all orifice jets with a single peak situated at lower drop diameters. Filament breakup is reasoned to be the cause of higher number of smaller drop sizes in triangular sprays among all non-circular orifice jets while thinner filaments in circular orifice sprays lead to smaller drop sizes compared to their triangular counterparts showing the same breakup morphology. We expect our results to help formulation of atomization strategies based on only changing the orifice geometry which heretofore have been achieved by complicated means thereby beneficially impacting applications as diverse as engine fuel combustion, pharmaceutical sprays and $CO_2$ capture by NaOH sprays.

## 1. Introduction

The atomization of liquid jets from orifices is of practical relevance in several practical applications such as aerospace propulsion, internal combustion engines, agricultural sewage and irrigation, powder metallurgy, manufacturing, biomedical and pharmaceuticals (Akinyemi et al. 2023; Ehteram et al. 2013; Jiayu et al. 2018; Kayahan et al. 2022; Lefebvre and McDonnel 2017; Makhnenko et al. 2021; Van Strien et al. 2022). In typical sprays, the bulk liquid is injected in the form of jets/sheets and transforms into fine, small sized droplets upon fragmentation (Lefebvre and McDonnel 2017; Sou A., Hosokawa S., and Tomiyama A. 2007). For instance, prior to the start of combustion, fuel sprays are converted to atomized droplets which aid fuel combustion. Stringent design requirements require an optimum exhaust gas temperature profile at the combustor exit and are greatly influenced by atomization characteristics of the burning fuel (Hiroyasu and Arai 1990; Jia et al. 2023). In other examples, such as inhalers (Longest and Landon 2012), sprinklers in irrigation systems (Jiang et al. 2019), $CO_2$ capture from atmosphere (Cho et al. 2018) specific drop sizes are desired for efficient, targeted delivery of pharmaceutical drugs, energy efficient water discharge and maximum $CO_2$ absorption respectively. Atomization of a liquid jet as enunciated thus far, is governed by a competition between inertia, viscous, and surface tension forces of the jet. These can be succinctly represented by three non-dimensional numbers, namely Reynolds number, $Re_l$, liquid Weber number, $We_l$, and Ohnesorge number, $Oh$, (Kulkarni and Sojka 2014; Lefebvre and McDonnel 2017) which are expressed mathematically as,

$$Re_l = \frac{\rho_l U_o D_o}{\mu_l}, We_l = \frac{\rho_l U_o^2 D_o}{\sigma} \text{ and } Oh = \frac{\mu_l}{\sqrt{\rho_l D_o \sigma}} \tag{1}$$

where $U_o$ is the jet velocity, $D_o$ is the jet diameter, $\rho_l$ is the density of jetting liquid, $\mu_l$ is the dynamic viscosity of the jetting liquid, and $\sigma$ is the surface tension (Dumouchel 2008). For a given $Oh$ (jet diameter and jetting liquid), the Ohnesorge's chart shows four jet breakup regimes for the liquid jet with varying $Re_l$: Rayleigh regime, first wind-induced regime, second wind-induced regime, and atomization regime (Lefebvre and McDonnel 2017). The present work focuses on the characteristics of liquid jets in the atomization regime, which typically occurs at high $Re_l$ (or $We_l$). For our analysis in this text we choose $We_l$ as the dimensionless number to describe our results due to its congruence to $Re_l$ and ability to describe surface tension effect more elaborately.

In the atomization regime, the jet is fully transformed into droplets and the average droplet size is much less than the jet diameter at the orifice exit unlike at low $We_l$. Since all applications stated previously either aim to operate in this regime or avoid it completely, the understanding of the mechanisms leading to atomization in this regime are extremely important with the orifice shape playing an important role. Most commonly, plain orifice atomizers designed with circular orifices are used in IC engines, aircraft jet engines (Nurick 1976), rocket thrust chambers, spray towers for carbon capture (Cho et al. 2018) , and in jet impact on a rotating packed bed (Xu et al. 2019). They are also seen in engine combustors involving jet-in-cross flow (for example, afterburner and scramjet combustors). However, non-circularity in the orifice shape to improve atomization quality is garnering attention in current research investigations. Examples of non-circular orifice geometries can be seen in recent designs of impinging type atomizers used in rocket engines, MEMS (micro electromechanical systems) based fuel injectors (Baik S., Blanchard J.P., and Corradini M.L. 2003), and diesel injectors (Chen et al. 2022; Hiroyasu and Arai 1990; Jia et al. 2023; Yin et al. 2020; Yu S. et al. 2018) .

The shape of a non-circular orifice is one of the parameters to influence the flow characteristics of liquid jet which manifests itself by inducing waviness in liquid jets discharging at low $We_l$. In a landmark study, Rayleigh,1879 (Rayleigh 1879) analyzed such undulations which appear mainly as interfacial capillary oscillations on liquid jets issuing from non-circular orifices and reported theoretical predictions on the wavelength of the oscillations in terms of the flow parameters and number of undulations. In elliptic jets such interfacial oscillations have been referred to as axis-switching and are known to affect jet instability and breakup characteristics greatly (Amini et al. 2014; Amini and Dolatabadi 2011, 2012; Bechtel S.E., Lin K.J., and Forest M.G. 1988; Gu, Wang, and Hung 2017; Jaberi and Tadjfar 2020; Kasyap, Sivakumar D., and Raghunandan B.N. 2009; Rayleigh 1879)

In contrast to the above, the characteristics of high $We_l$ jets from non-circular orifices in the context of fuel atomization in diesel injection systems have received little or no attention in literature (Farvardin and Dolatabadi 2013; Gu et al. 2017; Jaberi and Tadjfar 2020; Morad, Nasiri, and Amini 2020; Yin et al. 2020; Yu S. et al. 2018, 2019, 2021). Their study is especially warranted since limited research has indicated their superior spray characteristics as evidenced by faster breakup (Amini et al. 2014; Amini and Dolatabadi 2011, 2012), higher level of initial turbulence (Kasyap et al. 2009; Sivakumar et al. 2015), higher spray cone angle (Muthukumaran and Vaidyanathan 2014; Sharma P and Fang T. 2014, 2015), higher air entrainment rate (Bifeng et al. 2022; Farvardin and Dolatabadi 2013; Triballier, Dumouchel, and Cousin 2003; Yin et al. 2020), and lower liquid-phase penetration (Bifeng et al. 2022; Yu S. et al. 2018). However, most of these studies have employed orifices which are elliptical in cross section(Yu S et al. 2018), and the number of works dealing with other non-circular orifices such as rectangular, square, and triangular orifices in the atomization regime are limited (Kasyap et al. 2009; Kasyap T.V., Sivakumar D., and Raghunandan B.N. 2008; Sharma P and Fang T. 2014; Wang F and Fang T. 2015; Yu et al. 2019).

Among all spray characteristics the drop size distribution of spray droplets is the most important to determine the quality of liquid atomization, playing a critical role in understanding various processes as outlined above. Research on the droplet characteristics of liquid jets from non-circular orifices is particularly deficient with the role of orifice shape (elliptical, triangular, and square) not investigated adequately (Bechtel S.E. et al. 1988; Chen et al. 2022; Jaberi and Tadjfar 2020; Yu et al. 2019; Yu S et al. 2018). Notable among these studies are those by (Wang F and Fang T. 2015) who mainly reported on breakup length at low injection pressures of 5 bar for water jets of non-circular orifices. They found that axis-switching and increased aspect-ratio effect in rectangular jets suppresses the increase of breakup-length at these low pressures. On the other hand, square and triangular jets being affected by the ambient air are more unstable, especially at higher pressures, leading to shorter breakup-length compared to the circular and rectangular jets. These studies were followed by (Sharma P and Fang T. 2014) who reported that non-circular orifice water jets have shorter breakup lengths due to enhanced instability brought about by axis-switching for injection pressures up to 69 bar. Since neither of these studies considered actual fuels or drop sizes characteristics, (Sharma P and Fang T. 2015) investigated diesel sprays at much higher injection pressures of 300 to 1000 bar. They observed higher mixing at these pressures and also found that closer to the nozzle, non-circular sprays produce smaller droplets for injection pressures up to 500 bar whereas at higher pressures, droplets from circular sprays are smaller, a trend which is followed at larger distance as well.

In our work we aim to study the spray characteristics of Jet -A1 at injection pressures of 0.1-15 bar which has not been previously investigated in detail. Jet A-1 differs from diesel both in chemical composition and its fluid characteristics. Given that droplet characteristics in the atomization regime of liquid jets is crucial for the development of fuel injectors employing non-circular orifices, our present work aims to address this important gap in literature.

Specifically, we design two objectives of our current study, *first* we record emerging spray morphologies at different $We_l$ for liquid jets discharging from elliptical, equilateral triangular, and square orifices of same area of cross section in the atomization regime using Jet A-1 fuel as the jetting liquid. *Second*, we seek to understand the role of orifice shape on the droplet sizes resulting from the atomization of jets emanating from non-circular orifices and compare them with data on droplet sizes obtained on atomization of jets issuing from circular orifices. We connect our visualization from our first objective to the second thereby providing a comprehensive picture of atomization of jets emanating from non-circular orifices in the atomization regime. Our paper is organized as follows: we begin by providing details of our experiments followed by a discussion on the spray structure as inferred from shadowgraphy images after which we provide details on experimental drop size measurements.

## 2. Experimental apparatus and measurement details

To accomplish our objectives mentioned above we fabricate one circular and four non-circular orifices after which we install them in a spray test facility which discharges Jet-A1 (kerosene) fuel at different injection velocities. Details of all these are presented below.

### 2.1 Fabrication and geometric details of circular and non-circular orifices

Three non-circular orifices of different shapes, ellipse, equilateral triangle, and square, of equal area of cross section are considered in the study. The elliptical, triangular, and square orifices are referred to in the current work as EL, TR, and SQ, respectively. The orifices are made of stainless steel (SS-304) and are fabricated using electric-discharge machining (EDM). Figure 1 shows high resolution images of exit planes of the non-circular orifices, and the geometrical dimensions of the orifices are measured using these images.

The equivalent diameter, $D_{eq}$ of a circular orifice with the same exit cross sectional, $A_0$ as the non-circular orifice, is estimated as

$$D_{eq} = \sqrt{\frac{4A_o}{\pi}},  \tag{1}$$

In previous works [9], the hydraulic diameter, $D_h$ of non-circular orifice is used to normalize

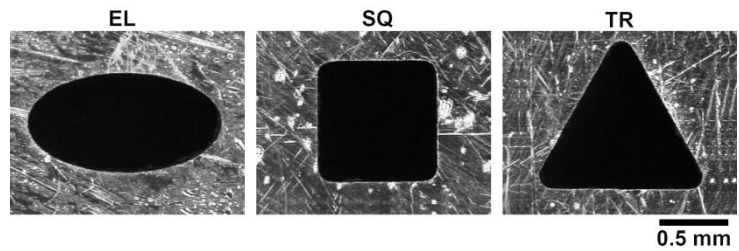

**Figure 1.** High-resolution images of exit planes (left to right) of the elliptical (EL), square (SQ), and triangular (TR) orifices.

Sauter mean diameter (SMD) of the spray, and it is calculated as

$$D_h = \frac{4A_o}{P_o},  \tag{2}$$

where $P_o$ is the wetted-perimeter of the orifice. An equivalent circular orifice, identified in the current work as CR, is also fabricated. The orifice length is kept at 5 mm for the fabricated value considered in previous studies [32]. Table 1 shows the salient geometric details of the orifices, which provides an orifice length-to-diameter (equivalent diameter) ratio closer to the non-circular and circular orifices.

| Orifice Shape | $A_o$ (mm$^2$) | Geometrical dimensions | $D_{eq}$ (mm) | $D_h$ (mm) |
|---|---|---|---|---|
| EL | 0.766 | Major axis length = 1.38 mm | 0.99 | 0.89 |
|  |  | Minor axis length = 0.71 mm |  |  |
| TR | 0.714 | Side length = 1.28 mm | 0.95 | 0.74 |
| SQ | 0.723 | Side length = 0.85 mm | 0.96 | 0.85 |
| CR | 0.882 | Diameter = 1.06 mm | 1.06 | 1.06 |

**Table 1** Geometrical details of non-circular circular orifices.

The fabricated orifice plate is arranged in an injector holder assembly (not shown here). The injector holder consists of a cylindrical chamber (with inner diameter = 15 mm and length = 15 mm), a diverging portion of circular cross section, and a circular tube (with inner diameter = 4.6 mm and length = 12 mm). The orifice plate is connected at the end of the cylindrical chamber by means of four M4 screws and the circular tube is connected to the liquid feed line.

### 2.2 Spray test facility, liquid, and flow properties

A schematic of the spray test facility used (Sivakumar et al. 2015; Sivakumar D. et al. 2016) in the present study is shown in Figure 2. A large stainless-steel tank is used to store the fuel under pressure by means of a compressed nitrogen gas supply and a pressure regulator. A filter with the nominal pore size 40 μm (particulate inline filter from Swagelok, USA) is employed between the tank and the orifice assembly to arrest contaminants in the flow line. The liquid flow rate, $m$ from the orifice is varied using a flow control valve and a pressure gauge arrangement and is measured using a flow meter (micromotion R series Coriolis type flow meter from Emerson, USA) with a measurement accuracy of 0.75%.

For a given test condition, the pressure drop across the orifice, $\Delta P$, and $m$ are noted from the test facility. The average velocity of liquid jet at the orifice exit, $U_o$ is estimated from $m$ and $A_o$ as

$$U_o = \frac{m}{\rho_l A_o}. \tag{3}$$

The flow experiments are carried out using Jet A-1 (aviation kerosene) fuel. The fuel is procured from Indian Oil Corporation, and whose properties (density, $\rho_l = 786.9$ kg/m³, dynamic viscosity, $\mu_l = 1.18 \times 10^{-3}$ Ns/m², and surface tension, $\sigma = 0.0255$ N/m) are determined in laboratory.

### 2.3 Measurement details for shadowgraphy and drop size characterization

The morphological characteristics of fuel sprays at different flow conditions from the non-circular and circular orifices are obtained from the analysis of digital images of the sprays captured via photographic techniques as followed in prior studies (Kulkarni et al. 2010; Sivakumar et al. 2015; Sivakumar and Kulkarni 2011). A Nikon D7000 digital camera fitted with a zoom lens (AF Zoom Nikkor 80–200 mm f/2.8D) and a diffused backlighting system are

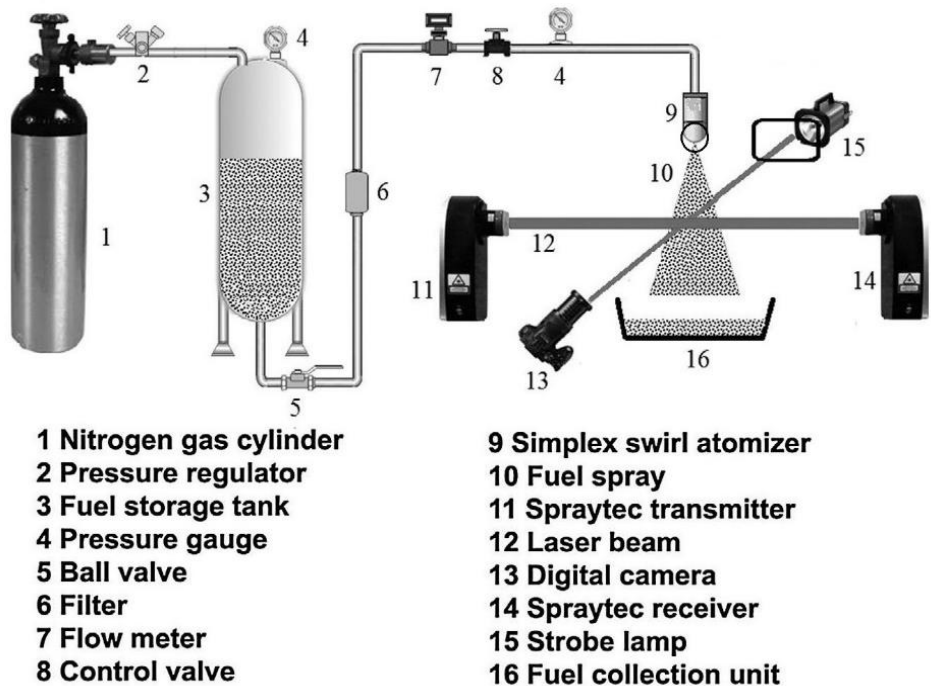

**1 Nitrogen gas cylinder**
**2 Pressure regulator**
**3 Fuel storage tank**
**4 Pressure gauge**
**5 Ball valve**
**6 Filter**
**7 Flow meter**
**8 Control valve**

**9 Simplex swirl atomizer**
**10 Fuel spray**
**11 Spraytec transmitter**
**12 Laser beam**
**13 Digital camera**
**14 Spraytec receiver**
**15 Strobe lamp**
**16 Fuel collection unit**

**Figure 2.** Schematic of the spray test facility with individual components numbered 1 to 16.

used to take capture high resolutions photographs of the sprays. The pixel resolution of the camera is $4928 \times 3264$, and images of the sprays are captured with the resolution of 8 pixels/mm. The captured digital images of the fuel sprays are analyzed by means of the image processing software ImageJ (Abramaoff, Magalhaes, and Ram 2004).

Laser diffraction based Spraytec equipment (Malvern Instruments, UK) is used to measure droplet size distribution of sprays from the non-circular and circular orifices at different axial distances from the orifice exit. The equipment provides a line-of-sight average measurement of the volume-based drop-size distribution from an analysis of the diffraction pattern resulting from the interaction between a test spray and a laser beam. Figure 2 shows the arrangement of Spraytec in the spray test facility (Vankeswaram et al. 2023). The equipment consists of a light transmitter unit with He-Ne laser as the light source, and a light receiver unit, which houses a collecting lens and a series of 33 diodes. The collecting lens has a focal length of 750 mm with the measurable droplet diameter in the range of $0.1 - 2500 \ \mu m$. The present equipment is used earlier to measure drop size distribution of fuel sprays, and additional operational details of the equipment operation are given in our previous works (Sivakumar et al. 2015; Sivakumar D. et al. 2016). The equipment provides accurate spray drop size measurements over a wide range of light transmission values (Triballier et al. 2003). The influence of multiple light scattering on the measurement of droplet size distribution is negligible if the light transition value is above 40%. For the measurements obtained in the present study, the light transmission value is recorded in the range 39% to 70%, and lower light transmission values are recorded for sprays with higher $\Delta P$. Lastly, the beam diameter is 20 mm which basically limits the radial measurements using this system for a liquid jet and hence we have taken only measurements at different axial locations downstream of the nozzle.

## 2.4 Experimental test conditions and correspondence to atomization regime

In our test conditions, the range of $We_l$ for circular and non-circular orifices vary from about 13023-14198 at the lower end to a maximum value of 46259-48510 with $\Delta P$ ranging from 0.55 – 1.69 MPa. The chosen $We_l$ range ensures a high level of turbulence in the jet (Sallam K.A., Dai Z., and Faeth G.M. 2002) exiting the orifice. These values are tabulated in Table 2.

| Orifice geometry | Range of exit velocity $U_o$ (m/s) | Range of $Re$ | Range of $We_l$ |
|:---:|:---:|:---:|:---:|
| **CR** | 20.8 – 38.5 | 14725 - 27218 | 14198 - 48510 |
| **EL** | 20.8 – 39.3 | 13720 - 25910 | 13224 - 47163 |
| **TR** | 21.9 – 40.5 | 13929 - 25764 | 14117 - 48293 |
| **SQ** | 21.0 – 39.5 | 13414 - 25281 | 13023 - 46259 |

**Table 2** Range of jet exit velocity, $Re$ and $We_l$.

In terms of the Ohnesorge chart, represented in terms of $Oh$ versus $Re_l$ (or equivalently for $We_l$) for the circular liquid jets (Lefebvre and McDonnel 2017) the liquid jets discharging from the different non-circular orifices with an equivalent circular jet diameter (see Table 1) such as those studied correspond to complete atomization at high $Re_l$ as depicted in Figure 3.

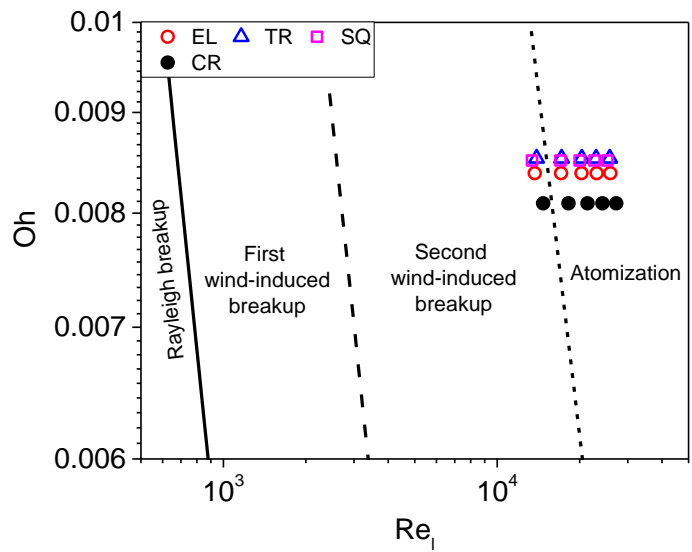

**Figure 3**: Ohnesorge chart highlighting different regimes of circular liquid jet breakup [1]. The symbols represent the current test conditions for fuel jets from the non-circular and circular orifices.

## 3. Results and discussion

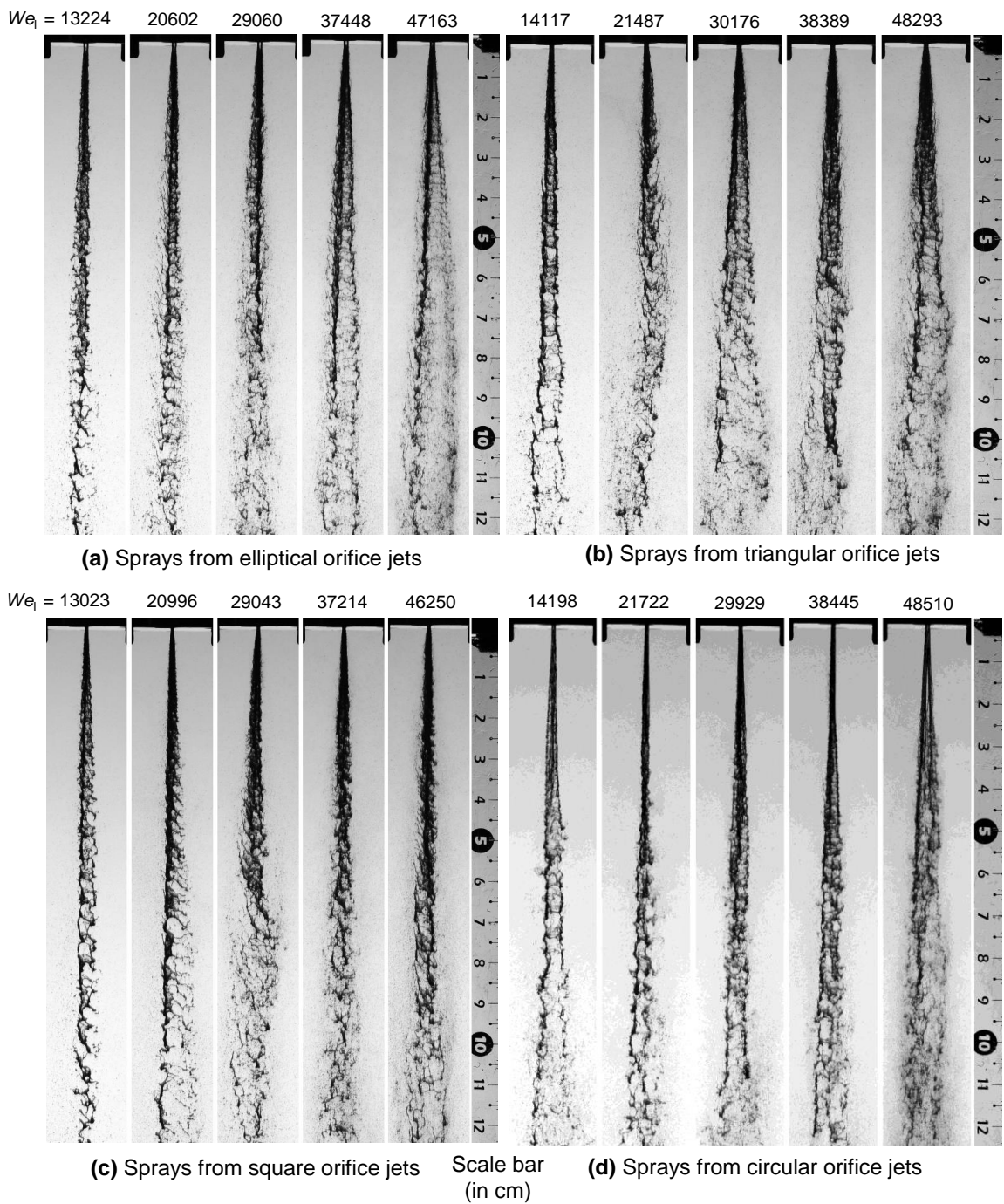

**(a)** Sprays from elliptical orifice jets     **(b)** Sprays from triangular orifice jets

**(c)** Sprays from square orifice jets     Scale bar (in cm)     **(d)** Sprays from circular orifice jets

**Figure 4**: Jets exiting orifices of different geometries at different $We_l$ corresponding to increasing injection pressures (a) Elliptic, (b) Triangular, (c) Square and, (d) Circular.

From our experiments we gather detailed visualizations through shadowgraphy (see Figure 4) which are further supported by drop size measurements at different high and low $We_l$. In this section we systematically analyze our data starting with describing the overall spray features and using them to understand the experimentally measured drop size characteristics.

## 3.1 Overall spray morphology and dominant features

Figure 4 (a)-(d) shows the overall behavior of Jet A-1 jet from different orifice configurations for the high (46259-48510) and low $We_l$ (13023-14198) conditions. Closely observing the structure of the spray at these conditions we note distinct behavior at low $We_l$ for jets emanating from, (*i*) square and elliptic, compared to (*ii*) circular and triangular orifices. At high $We_l$ this behavior is dominantly influenced by the high inertia and a near similar behavior is seen for all orifices.

Our further discussion is organized along these lines wherein we first discuss the overall behavior at low $We_l$ and then move to a general description of spray behavior at high $We_l$ finally describing spray features at location much downstream when the spray is fully atomized.

### 3.1.1 *Behavior at low $We_l$*

From our visualizations as shown in Figure 4 (a)-(d) we identify two distinct morphologies – (i) filament-sheet breakup, seen in circular and triangular orifice sprays and, (ii) core breakup seen in elliptic and square orifice sprays as shown in detail in Figure 5(a) and (b).

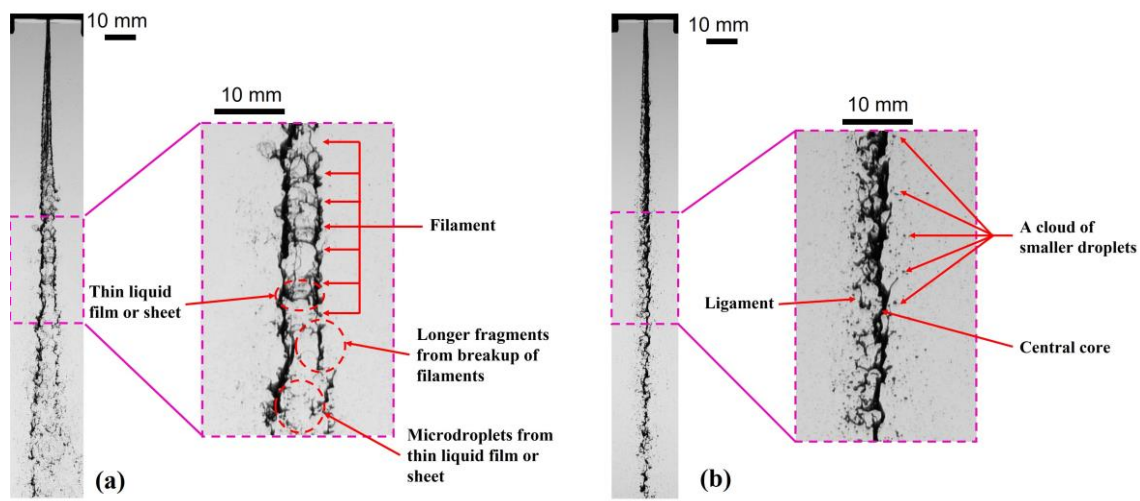

**Figure 5 (a)**: Filament breakup features as seen in circular and triangular orifices. The main elements are two bounding filaments and a thin liquid film. Filament breakup into large drops while thin film bursting leads to microdroplet generation **(b)** Core breakup features as observed for elliptic and square orifice. The major features are a central core which give rise to ligaments which generate drops of a large size and is surrounded by a cloud of smaller droplets resulting from surface atomization.

- *Filament-sheet breakup* : Upon exiting, jets from circular and triangular orifices expand as they move downstream in the ambient atmosphere displaying two main features – filament and a thin film bounded by this filament as seen in Figure 5(a). This expansion leads to a widening of the spray and a decrease in average velocity of the intact region of the spray with a gradient of velocity which increases from center outwards (Castrejon-Pita, Hoath, and Hutchings 2012) thereby effectively conserving mass. This implies that as the jet moves downstream, and as the effect of inertia decreases the influence of the surface tension force on the edges of the jet increases resulting in the formation of a bounding filament which encloses a thin liquid membrane and is described schematically in Figure 6(a). Note that for triangular orifices we expect the jet to spray to first relax to circular cross-sectional profile (Middleman 1964) and then follow exhibit the same behavior as the jet from a circular orifice.  Therefore, we mainly see two features specific to low $We_l$ circular and triangular jets: filaments and thin liquid films (or sheets) contained between the filaments(Chen and Ashgriz 2022). The filaments eventually transform to ligaments (Kinnersley and Parkin 1995; Raza and Sallam 2023)which the thin liquid film ruptures to form microdroplets (see Figure 5(a)).

- *Core breakup* : For square and elliptic orifices, the issuing jet is seen to be devoid of any filaments and instead comprises of a core which is surrounded by cloud of surface atomized microdroplets as shown Figure 5(b). The core further disintegrates in ligaments producing large droplets(Chen and Ashgriz 2022). To rationalize this observation, we see that for jets discharging from square and elliptic orifices we see that as the jet exits the orifice it displays an almost uniform plug velocity profile across its width. This is so because of the 2D nature of the exiting spray which closely follows the structure of a liquid sheet which for the same $We_l$ and cross-sectional area of a circular or triangular jet must conserve mass and therefore attain a velocity profile which is much flatter. Due to this there is no mass accumulation in the bounding filament and the spray breaks up without the formation of a filament, which we term as core breakup shown schematically in Figure 6(b). Instead in such a breakup we primarily see formation of ligaments which break up into drops of large sizes and are shrouded by a mist of fine droplets due to surface atomization as will be described below.

### 3.1.2 *Behavior at high $We_l$*

At higher $We_l$ the formation of filament is not dominantly seen for circular and triangular orifices since at higher velocities the effect of surface tension is less dominant compared to inertia of the liquid jet (Kasyap T.V. et al. 2008)(Bechtel 1989). This is further supported by the fact that the velocity profile evens out to a flatter profile for both square and elliptic orifices at higher $We_l$ as turbulence levels increase in the liquid jet. Similarly, for jet discharging from square and elliptic orifices we dominantly see breakup of a core liquid structure into ligaments. Surface atomization due to high shear between the liquid and the air is also accentuated at these conditions. More on this shall be discussed in the section on drop size measurements.

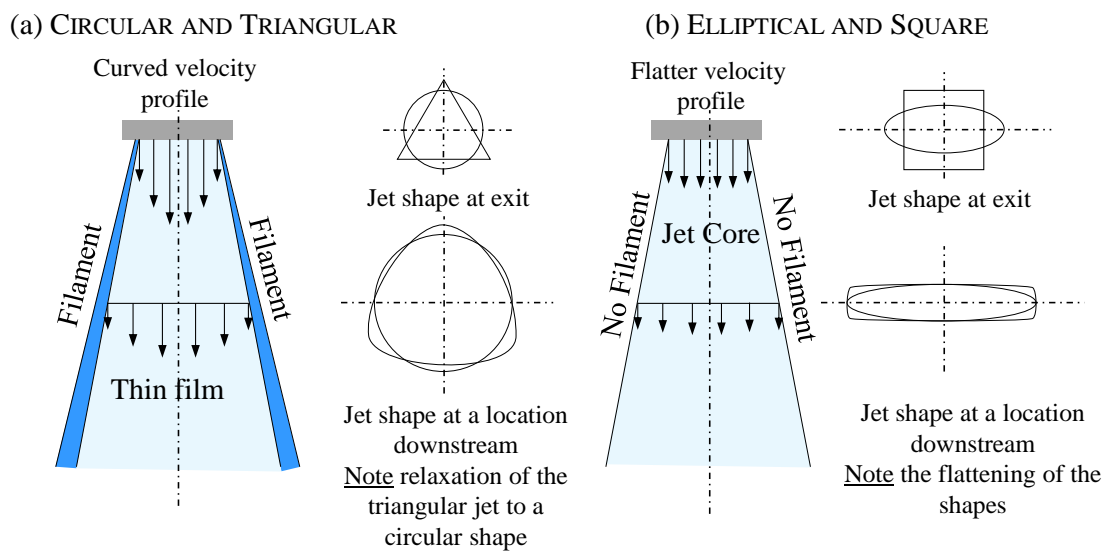

**Fig 6(a)** Hypothesized mechanism of filament breakup in circular and triangular orifice liquid jets. A filament is formed due to lower velocity at the edge of jet implying higher influence of surface tension. For circular jets the surface tension force is lower compared to triangular jets as the orifice exit curvature for a circular jet is lower. This causes the triangular jet filaments to become thicker. **6(b)** Mechanism of core breakup in elliptic and square orifice liquid jets. A more uniform exit velocity profile ensures that inertia is high at the edges of the spray and surface tension is not as dominant. Surface atomization is predominantly seen at the edges here while the core undergoes breakup by itself often resulting in large ligaments which breakup in large drops. Envelopes of the jet spray cross section shown alongside the figures **(a)** and **(b)**.

## 3.2 Spray morphology at distances far downstream of the nozzle

In this section we describe the consequence of the observed spray morphology at distances far downstream of the nozzle, around 15-20 cm from the orifice exit where the spray transits from primary breakup to secondary breakup regime. Here we specifically discuss the fate of three liquid jet structures (i) ligaments, (ii) filaments, (iii) thin liquid films and, (iv) surface atomization which form the constituent elements of breakup features once the jet disintegrates and encompass observations of spray features at all $We_l$.

### 3.2.1 *Breakup of ligaments and filaments*

With the increase in distance from the orifice exit, the fuel ligaments and filaments (upon detachment) identified in both filament and core breakup previously continue to transform and eventually form droplets. The images given in Figure 7 highlight the flow of ligaments inside the spray of fine droplets. The breakup of these ligaments/filaments results in the production of several big droplets in the spray region which is seen in Figure 7 as the production of large droplets and some smaller satellite droplets in the spray. This means that the large droplets in the spray can be correlated to the size of the ligaments or detached filaments seen in the images. It is also evident that, among the non-circular orifices (Figures 7(b), 7(c), and 7(d)), the elliptical orifice produces highly interconnected ligaments, and the presence of these ligaments is not widely distributed appearing within a small radial distance of each other. On the contrary, the triangular orifice produces highly separated ligaments, which are distributed over a larger radial distance. The ligaments formed by the CR orifice (Figure 7(a)) are significantly thinner than that from the non-orifices (Figs. 7(b) – 7(d)), which suggests better atomization with the circular liquid jets.

At higher $We_l$, the jet is more unstable which results in thinner ligaments and faster ligament breakup at higher $We_l$. While ligaments are still present in the spray for the fuel jet from the EL orifice (Figure 7(b), the spray from the TR orifice (Figure 7(c)) does not contain almost any ligaments. A finely dispersed spray is observed for the fuel jet from the CR orifice (Figure 7(a)). Figures 7 (a) and (b) show qualitatively that the orifice shape influences the atomization of the fuel jet from the circular and non-circular orifices.

A detailed description of ligament and detached filament stability dependent on their initial shape is elucidated in appendices 1 and 2 wherein these are characterized in terms of $We_l$ and orifice shape.

### 3.2.2 *Breakup of thin liquid film and surface atomization*

In the case of circular and triangular orifice where filament breakup is observed, a thin liquid film contained between the filaments is seen which is O (few μm) which eventually disintegrates into fine droplets when it ruptures. This leads to formation of microdroplets, reducing the overall diameter of the drop size at that specific location. Another mechanism by which such fine droplets can be produced is surface atomization which typically occurs when the surface of the jet shears against the ambient air leading to formation of small wavelength Kelvin-Helmholtz waves which produces small droplets continually ejecting circumferentially from the jet(Beale and Reitz 1999; Varga, Lahseras, and Hopfinger 2003). Expectedly with increasing jet velocity this behavior is pronounced leading to finer droplets as seen in Figure 7 (a)-(d). Sections 3.2.1 and 3.2.2 therefore aim to provide a description of the fate of the finely atomized and coarsely atomized droplet resulting from primary breakup. In our following

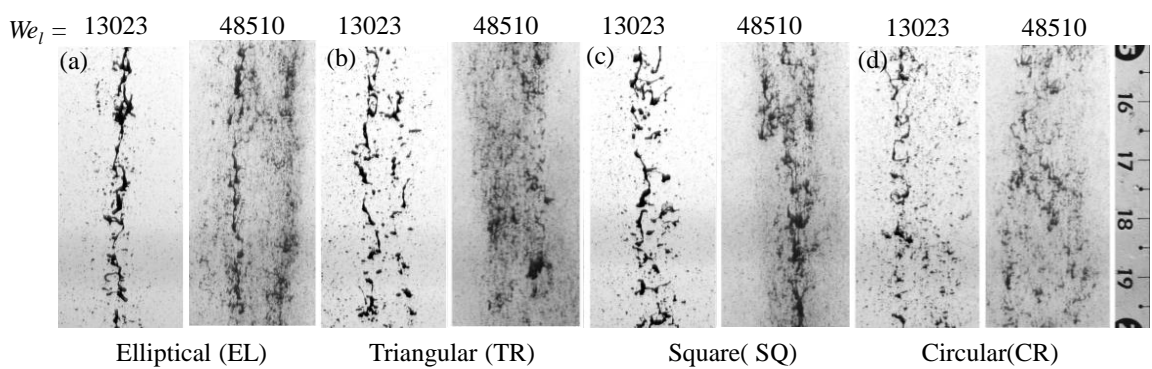

**Fig. 7.** The spray morphology at distance $15 - 20$ cm from the orifice exit for the Jet A-1 jet for the maximum and minimum $We_l$ [13023 and 48510] from different orifice configurations, **(a)** EL, **(b)** TR, **(c)** SQ, and **(d)** CR. Scale bar on right is in cm.

discussion we establish the connection of these spray breakup features to the observed drop sizes.

### 3.3 Droplet size characteristics

Here we present the size characteristics of spray droplets formed from the Jet A-1 jet with different flow conditions from the circular and non-circular orifices. We begin by describing the mean drop sizes followed by droplet size distribution is measured using Spraytec, and for a given flow condition, the measurements are obtained with focus on specific axial locations from the orifice exit, $Z$. First is the location where the drop size is a maximum usually associated with the end of primary breakup and, second and last is the location far downstream where secondary breakup commences. This helps us provide an overall picture of the jet and drop atomization.

### 3.3.1 *Mean drop sizes produced from circular and non-circular orifice sprays*

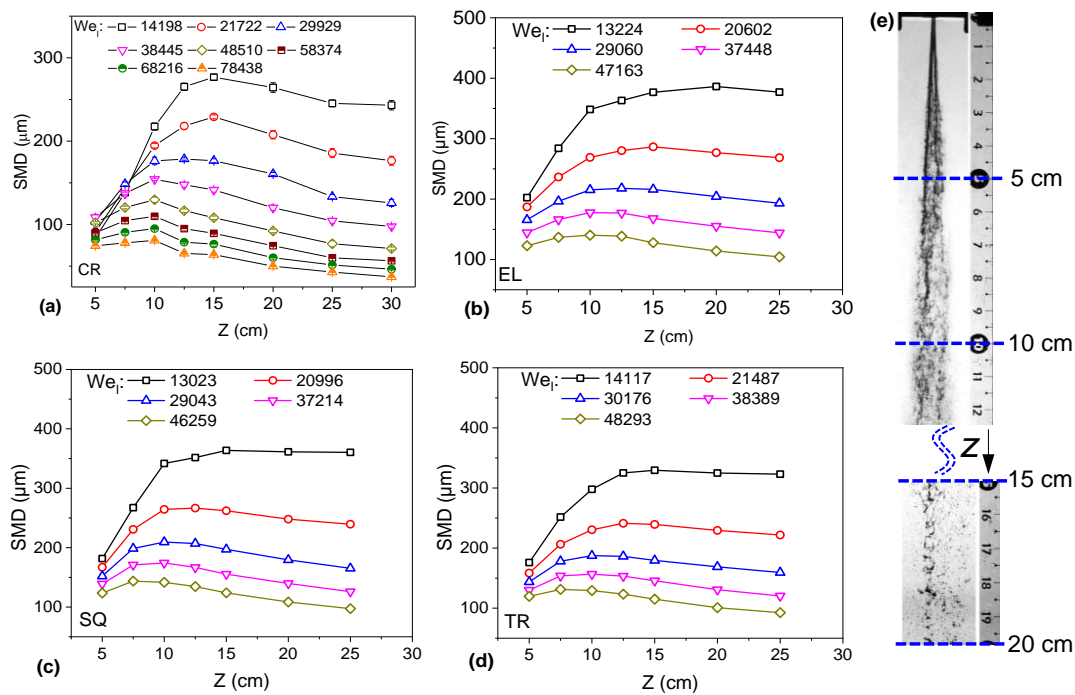

**Figure 8.** The variation of SMD with $Z$ (*in cm*) for the Jet A-1 jet at different $We_l$ for (a) circular (CR) (b) elliptical (EL) (c) square (SQ) and, (d) triangular (TR) orifices, (e) Shadowgraphy image showing $z$ locations at which drop size measurements are made.

A spray consists of a range of drop sizes describing which can be difficult especially if we want to compare quality of atomization at different conditions quickly(Kumar et al. 2019). To enable an easy yet informative comparison computing the mean diameter of an ensemble of spray droplets at a given condition can often be helpful(Hewitt 1993; Vankeswaram and Deivandren 2022). Here we use one such metric known as the Sauter Mean Diameter (SMD) or $D_{32}$ which is the diameter of a monodispersed spray having the same total volume to surface area ratio as that of a given spray.

(*i*) *Variation of mean drop size for Jet A-1 jet from circular and non-circular orifice*: Here we examine the variation of Sauter mean diameter, SMD with $Z$ as shown in Figure 8 (a) –(d) for the Jet A-1 jet from the circular and non-circular orifice at different $We_l$ corresponding to different $z$ locations (see Figure 8(e)). Each data point in the plot corresponds to an average of 5 measurements recorded at different instants.

- *At low $We_l$* the SMD increases sharply with increase in $Z$, reaching a maximum, and then decreases gradually with the increase in $Z$. To understand this better we refer to our previous discussion (Section 3.1) and Figure 5 where we see for circular jets a filament is formed which bounds a thin liquid film. At locations close to the injector exit this thin liquid film bursts producing drops of a small size. Moving further downstream we see that the filaments too disintegrate generating drops of larger size increasing the mean drop size and corresponds to the peak in Figure 8(a) for $14198 < We_l < 38445$. This peak is followed by a reduction in drop size at locations further downstream due to breakup of the ligaments generated at upstream locations. Further we also observe from the plot that the slope of the early SMD rise curve (at low values of $Z$) decreases with the increase in $We_l$. It is apparent from the images of Jet A-1 jets given in Figure 4 that, at $Z \sim 5$ cm – 12 cm (region close to injector exit), the atomization of the fuel jet is not fully complete, and the spray portion exhibits significant presence of jet core segments. The location of maximum SMD (SMD$_{max}$) seen in the plot can be safely considered as the location at which the bulk liquid transforms into smaller fragments (ligaments or drops) also known as primary atomization, reaches its completion.

Non-circular orifice jets (Figs. 8(b)-(d)) are marked by absence of a clear peak. This is due to surface atomization (for EL and SQ) and thin film breakup (for TR) that is relatively near the injector exit which produce small size drops. However, further away core breakup (for EL and SQ) dominates and leads to large sized drops. In the case of triangular orifice jets (TR) the breakup is similar to circular orifice jets with thicker filaments (as described in the next section) which displays similar behavior as EL and SQ orifice jets. The drops produced at location Z = 10 cm are relatively stable an do not undergo further breakup thereby extinguishing the peak seen for circular jets.

- *At high $We_l$* between 48510 and 78438, the filament diameter (for CR and TR) becomes thinner as the velocity profile become flatter and therefore, we see a depression in the peak (seen later in Figure 12). A similar trend is seen for EL and SQ orifice jets where core ligaments become thinner and the curves gradually flatten out.

To validate our current experimental observations with measurements of the mean drop sizes in circular sprays we compare our results with those from Hiroyasu and Arai, 2008 as shown in appendix 3 and observe a good agreement helping us extend our analysis for the Jet A-1 jet to non-circular orifices geometries.

As mentioned previously, of significance to us is the maximum value of SMD (denoted as $SMD_{max}$) for every $We_l$ among all $Z$ locations which shall be used a marker for the end of primary atomization and SMD measurements at Z = 25-30 cm where secondary atomization (Kulkarni and Sojka 2014) commences.

(*ii*) *Effect of orifice shape on mean drop size of non-circular jets*: To analyze the effect of orifice shape on the mean drop size of the Jet A-1 jet from the non-circular orifices we plot the variation of dimensionless $SMD_{max}$ (scaled by $D_{eq}$) with $We_l$ as shown in Figure 9. $SMD_{max}$ is obtained through the procedure described in the previous section, with $We_l$ for the fuel jet from the circular and non-circular orifices.

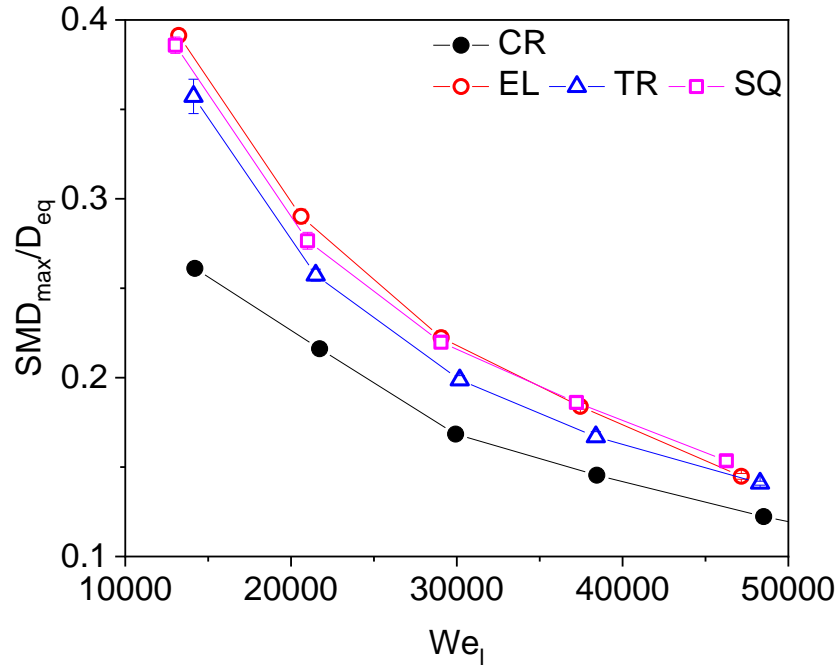

**Figure 9.** The variation of SMD$_{max}$ with $We_l$ for the Jet A-1 fuel jet from the circular and non-circular orifices.

On examining this in detail we notice that for a given $We_l$, the fuel jet from the CR orifice develops a smaller SMD$_{max}$, especially at low $We_l$ due to thin filaments and even thinner liquid films which burst producing microdroplets. Filament breakup due to this inherent morphology is expected to lead to better atomized drops viz-à-viz core breakup which involves disintegration of large masses of the liquid core with comparatively minor contribution from surface atomized drops.

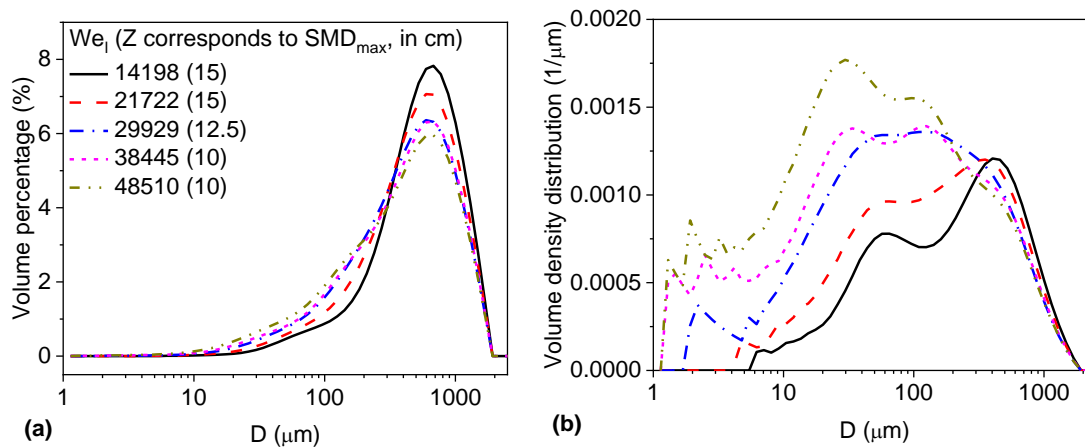

**Figure 10. (a)** The volume percentage and **(b)** volume density distribution curves correspond to location of SMD$_{max}$ of the Jet A-1 jet with different $We_l$ from the circular orifice plotted on a semi-log scale. The data below 10 μm is subject to lot of uncertainty, owing to limitations of the instrument and not considered in this analysis.

Therefore, among the non-circular orifices, the triangular orifice jets exhibit the better atomization characteristics in terms of $SMD_{max}$ compared to that of the elliptical and square orifices which primarily undergo core breakup. We can understand this by noting that triangular orifice jets show the same filament, thin liquid film breakup as circular orifices however they tend to contain thicker filaments than circular orifice as more liquid near the edge of the jet is influenced by surface tension due to higher curvature of the orifice. The ramifications of this observed behavior can be further obtained through the analysis of droplet size distribution which we consider in our ensuing discussion.

### 3.3.2 *Drop size distribution in circular jets measured at Z = Z at $SMD_{max}$*

Figure 10(a) and (b) shows the droplet size distribution curves for the Jet A-1 jet from the CR orifice at flow conditions corresponding to that shown in Figure 9 (filled black circles). Figure 10(a) shows the raw data given by the Spraytec in terms of volume percentage of spray droplets versus droplet diameter, $D$. In the figure, the flow conditions of the jets are highlighted in the legend in terms of $We$, and the axial distance from the orifice exit ($Z$) in cm corresponding to the measurement of $SMD_{max}$ at each flow condition is mentioned inside the parenthesis shown in the highlighted legends. A salient feature of this distribution is the peak in the size range of large droplets displayed for a given flow condition in volume percentage curve, Figure 10(a). With increase in $We_l$, the position of the peak in the droplet diameter scale is almost unchanged along with a marginal decrease in the volume percentage of the droplets. However, the volume percentage of small droplets exhibits a rise with the increase in $We_l$ (Figure 10(a)), which suggests a greater number of small droplets at higher flow condition. Since the droplet volume scales with $D^3$, in the absence of droplet count distribution, the $D$ value corresponding to the peak seen in Figure 10(a) cannot be taken as a measure of characteristic size of droplets produced just at the completion of the primary atomization of the liquid jet. The volume percentage curves produced by the Spraytec hide the details of small droplets in the sprays due to unequal bin widths configured in the measuring equipment (Chen S. and Ashgriz N. 2022).

Alternatively, the volume percentage curves can be converted to volume distribution curves in which the area under the curves sums to unity, as seen with the probability distribution curves. The volume distribution curves reveal more details about the small droplets in the spray, and to get volume distribution from volume percentage, the recorded volume percentage data needs to be divided by the corresponding bin width. Figure 10(b) shows the volume distribution curves corresponding to the volume percentage curves given in Figure 10(a).

For the low $We_l$ (= 14198) jet, the volume distribution curve shows a bimodal distribution with two peaks at $57 - 65$ μm and $403 - 459$ μm for small and large droplets, respectively. Referring to our discussion in Section 3.1 this can be attributed to the presence of both filaments and microdroplets emanating from the thin liquid filament. The droplet diameters corresponding to the first peak are due to the spray of microdroplets formed on bursting of thin liquid film bounded by the two filaments (see the images given in Figure 5(a) and (c)). An increase in $We_l$ moves the first peak towards a lower value of $D$. The second peak appearing at large droplet diameters corresponds to droplets formed from the breakup of thick filament bounding the thin liquid film. The characteristics of such thick ligaments are quantified in appendix 1. The droplet diameter corresponds to the second peak is comparable to the measured average value of the ligament diameter ($<D_c>$), particularly at low $We_l$ conditions (see appendices 1 and 2).

Further, for the low $We_l$ jet, the probability of the second peak at large droplets is higher than that of the first peak at small droplets, which suggests that the low $We_l$ spray is marked with a significant volume of large droplets. However, for high $We_l$ jet, the spray is marked with a significant volume of small droplets as seen by the higher probability of the first peak.

### 3.3.3 *Drop size distribution in non-circular jets measured at Z = Z at SMD$_{max}$*

Figure 11 shows the volume distribution curves of spray droplets, deduced from the data obtained using Spraytec, for the Jet A-1 jet from the three non-circular orifices for the high and low $We_l$ corresponding to that shown in Figure 9 (open symbols). As described above (and in Figure 8), the $Z$ corresponding to the measurement of SMD$_{max}$ at each $We_l$ is mentioned in the highlighted legends inside the parenthesis. For each $We_l$, the corresponding curve of the circular jet shown in Figures 11 (a) and (b) is also included in the comparison plot as black line.

In general, at low $We_l$ among the non-circular orifices, the size distribution of the Jet A-

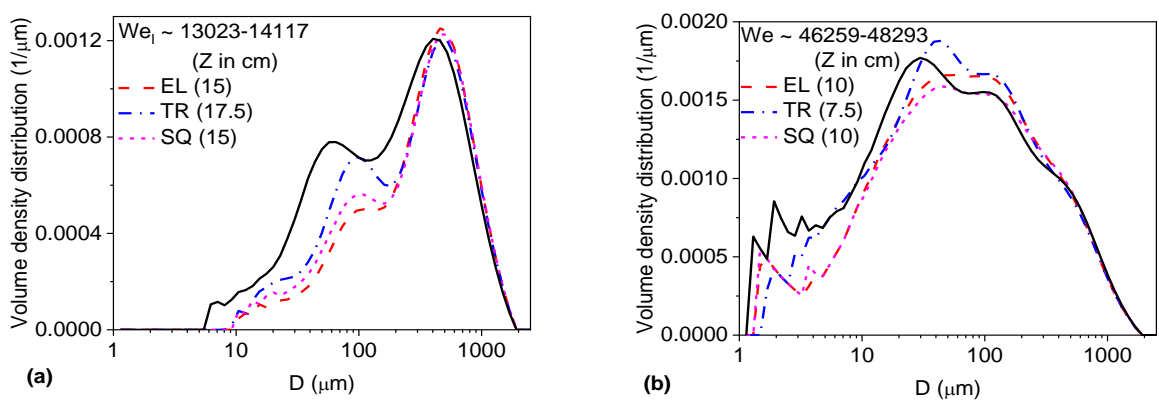

**Figure 11.** The volume distribution curves correspond to SMD$_{max}$ of the Jet A-1 jet for the (a) low $We_l$ and **(b)** high $We_l$ from the non-circular orifices. In each plot, the corresponding curve of the circular jet (given in Figure 10(b)) is shown as a black line.

1 spray from the triangular orifice (blue dash-dot curve) reveals the presence of higher volume PDF of small droplets and lower volume PDF of large droplets, irrespective of the flow condition. This spray behavior results in a lower SMD$_{max}$ for the Jet A-1 sprays from the TR orifice compared to that from the EL and SQ orifices (Figure 9). We can relate this to filament type breakup seen in triangular (and circular) jets which produces many smaller microdroplets due to film bursting compared to core breakup seen in elliptic and square orifice jets. This is even though smaller sized droplets are generated in core breakup due to simultaneous surface atomization, however they are not as dominantly present as the microdroplets from thin film bursting. The filament and thin film bursting process for triangular and circular orifices generates drops of two dominant sizes, first are large drops due to filament breakup and the second is the microdroplets from bursting of thin film.

In a similar manner, between elliptic and square orifice jets, the size distribution of the Jet A-1 spray from the EL orifice (red dashed line) reveals the presence of lower volume PDF of small droplets and higher volume PDF of large droplets, which results in higher SMD$_{max}$ for the Jet A-1 jets from the elliptical orifice (Figure 9).

At higher $We_l$ atomization quality becomes better due to increased inertia which is consistently observed across jets issuing from all orifices and consequently the peaks smoothen out.

**3.3.4** *Drop size characteristics at locations far downstream at Z = 25 and 30 cm*

We conclude our discussion with the characterization of spray droplets far downstream from the location of SMD$_{max}$ ($Z = 25$ and $30$ cm) to understand the drop size characteristics when secondary atomization begins to take effect.

- *Mean drop sizes for circular and non-circular orifices*: For a given $We_l$, the SMD of the circular Jet A-1 jet is significantly lower than that from the non-circular orifices. Among the non-circular orifices, the Jet A-1 jet from the elliptical orifice exhibits poor quality of atomization in terms of the mean drop size. The trend of SMD trend as shown in Figure 12(a) is like that seen at the end of the primary atomization plotted in Figure 9. For both for circular and non-circular there is only marginal decrease in the SMD at each $We_l$ compared to the values plotted in Figure 9. This seems to be at odds with the fact that secondary atomization leads to finer atomization. To understand this further we need to look at the drop size distribution in detail shown in Figures 12 (b) – (d) which is discussed in the next section.

- *Drop size distribution in circular and non-circular jets measured far downstream at Z = 25 and 30 cm* : On plotting the detailed drop size distribution for circular and non-circular orifice jets as shown in Figs. 12 (b), (c) and (d) we observe that the volume distribution curve becoming increasingly unimodal and is particularly apparent for high $We_l$ jets (see Figure 12 (d)). In other words, the effect of the ligament and filament diameter (the second peak) may be seen only at the primary atomization zone of the liquid jet. This further justifies that the second peak relating to the higher drop sizes is linked to the droplets formed from the thick filaments seen bounding the thin liquid film and from ligaments emanating from the core.

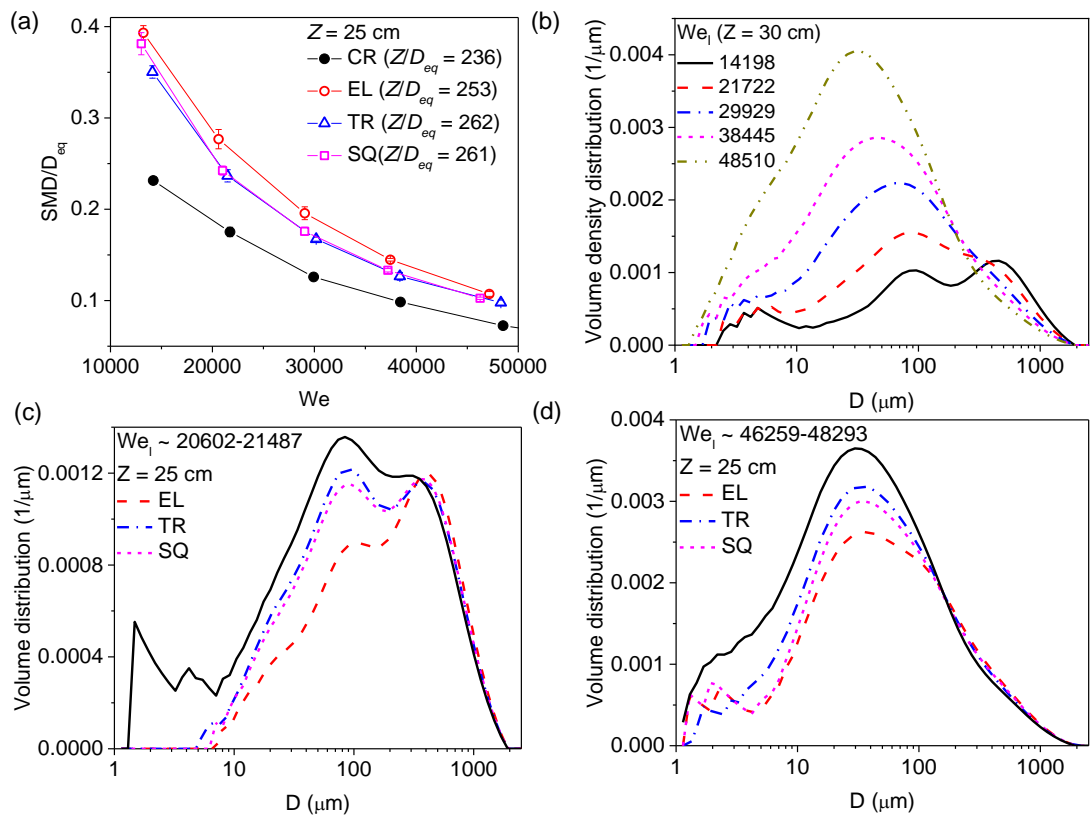

**Figure 12. (a)** The variation of SMD at $Z = 25$ cm with $We_l$ for the Jet A-1 fuel jet from the circular and non-circular orifices. **(b)** The volume distribution curve obtained at $Z = 30$ cm for the Jet A-1 jet with different $We_l$ from the CR orifice. **(c)** and **(d)** The effect of non-circular orifice shape on the volume distribution of spray droplets at $Z = 25$ cm from the orifice exit for the Jet A-1 jet with different $We = 20602$-$21487$ for **(c)** and, $46259$-$48293$ for (d) In each plot, the data of the circular jet is shown as black line.

To conclude we explain three key inferences from Figure 12 (a). First, the analysis of volume distribution curves of spray droplets at the downstream location of $Z = 25$ cm (Figures 12 (b)-(d)) reveals that the circular jet continues to produce more small droplets compared to atomized jets from non-circular orifices explaining the trends seen in Figure 12 (a). Second, the presence of significant volume of large droplets which have not undergone complete secondary atomization in the spray of the Jet A-1 jet from the elliptical orifice results in its higher SMD. Lastly, even though the bimodal behavior which consists of a peak at large diameter which increases the SMD is not seen at these larger *Z* locations, the unimodal distribution sees an increase in the volume of droplets in between the two peaks which can be deduced by comparing Figs. 12 (c), (d) and 11(a), (b) therefore the net effect on the mean drop size, SMD almost remains the same as noted in the beginning of this discussion.

## 4. Conclusions

The spray morphology and the drop size characteristics of Jet A-1 jets issuing from sharp-edged non-circular and circular orifices of comparable area of cross section are studied experimentally.

1. *Spray morphology*: We identified three dominant spray features: filaments, ligament and thin liquid film contained between the filaments besides surface atomization. Circular and triangular orifice jets are found to produce droplets via filament and thin liquid film bursting generating drops of large and small sizes respectively. Contrarily, square, and elliptic orifices do not exhibit such a morphology being dominated by breakup of the core which primarily produces ligaments with some evidence of surface atomization.

2. *Finer atomization of circular jets compared to non-circular orifice jets*: We find that circular orifice jets show the least mean drop size (SMD) at the end of primary breakup at all flow conditions due to thinner filaments and bursting of the thin liquid film bound by those filaments. Among the non-circular orifices, the triangular orifice jets produce smaller SMD compared to that of the elliptical and square orifices which we attribute to filament and thin liquid film bursting type breakup which produces finer atomization. A similar trend is seen at distances further away from the injector exit for drop sizes produced from jets from all orifices with a marginal decrease in the diameters due to secondary atomization.

3. *Bimodality in the probability distribution for circular and non-circular jet sprays*: In circular jet sprays these sizes are attributed to thin film bursting generated microdroplets and filament breakup produced large diameter sizes while in non-circular orifice jets the bimodality, though present, is less prominent as the core breakup via ligaments produced larger drops and smaller drops are mainly a result of surface atomization.

4. *Effect with increasing $We_l$ and large distances from the injector exit*: In both these cases the distribution becomes more unimodal for both circular and non-circular orifices, peaking at a lower diameter size due to the enhancement in liquid atomization by higher liquid inertia and increased jet instability for large $We_l$ and secondary atomization at large distances.

Finally, our results have huge implications considering the unique insights revealed through our investigations. For instance, non-circular orifice jets are known to produce shorter breakup length, higher instability which leads to finer drop sizes at low injection velocities corresponding to the axis switching regime (Amini and Dolatabadi 2011; Sharma P and Fang T. 2014). However, our results show that this behavior is not followed in the atomization regime circular orifice jets are found to be atomization more finely. From the perspective of combustion and applications such as drug delivery which require finer atomization the bimodality of spray jets from circular and triangular orifices may be undesirable. In $CO_2$ capture through NaOH sprays larger drop sizes may be required. So, the current work is expected to guide such studies with changing the orifice geometry as a one possible route to achieve expected drop sizes.

**Acknowledgements**

This work is carried out with the support of National Centre for Combustion Research and Development (NCCRD), Indian Institute of Science, India. One of the authors (SD) acknowledge the support from Department of Science & Technology (DST), India under FIST Program.

**Declaration of competing interest**

The authors declare that they have no known competing financial interests or personal relationships that could have appeared to influence the work reported in this paper.

**Appendix 1: Characteristics of ligaments and detached filaments produced from core and filament breakup**

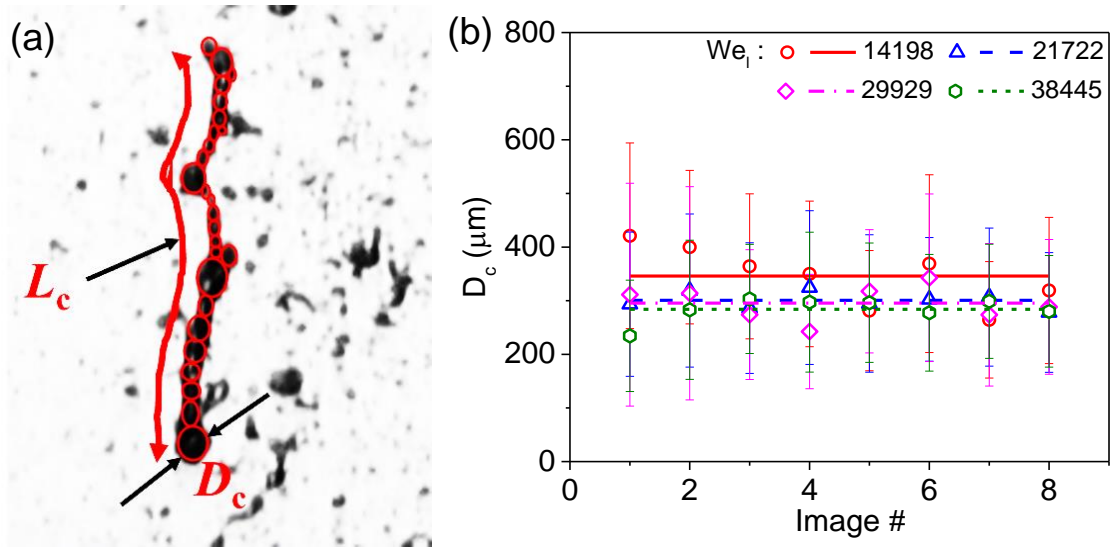

**Figure A1 (a)** A ligament/detached filament of Jet A-1 spray from the CR orifice with fitted blob circles, and the measurement details of the ligament/detached filament length **(b)** The value of $D_c$ measured from the images of Jet A-1 jet with different $We_l$ from the CR orifice. The lines represent the calculated value of $<D_c>$ at different $We_l$.

The size of droplets formed from the breakup of ligaments and detached filaments (referred just as filaments for brevity) is strongly correlated to the geometrical size of the ligament, particularly the ligament diameter (Marmottant P. and Villermaux E. 2004). The detached ligaments/filaments breakup due to capillary instability producing droplets to form a spray. For the Jet A-1 spray from the orifices, the ligament/filament diameter, $D_c$ is determined by measuring the diameter of the blob circles that can fit in the ligament/filament contour profile, as explained in ligament mediated fragmentation via an aggregation scenario (Eggers and Villermaux 2008; Marmottant P. and Villermaux E. 2004; Villermaux 2007; Villermaux, Marmottant, and Duplat 2004). For a given ligament, $D_c$ is taken as the calculated average of the diameter of all blob circles in the ligament (Figure A1(a)) and its length, $L_c$ is measured as highlighted in Figure A1(a). For each $We_l$ condition, a set of 15 images are used to determine the value of $D_c$ and $L_c$. In a single image, several ligaments/filaments are analyzed by fitting blobs inside their contours.

Figure A1(b) shows the variation of $D_c$ with image # (number assigned to images in a set of 8 images for a given flow condition) for a given flow condition of the jet from the CR orifice. From a single image, several ligaments/filaments are considered for the analysis, and the value of $D_c$ shown in Figure A1(b), A2(a) is the averaged value of $D_c$ of all ligaments/filaments recorded from the image. In general, the value of $D_c$ remains unchanged between the images. For a given jet condition ($We_l$), an average value of the ligament/filament diameter, referred to as $<D_c>$, is calculated from the recorded variation with the image #. The calculated value of $<D_c>$ for the Jet A-1 jet from the CR orifice is shown in Figure A2 (a) as lines for different $We_l$. According to Shinjo and Umemura (Shinjo and Umemura 2010), when the local gas Weber number of the ligament, $We_g$, expressed as $We_g = \rho_g U_r^2 a_L/\sigma$, is of the order unity ligament/filament may be observed. Here $U_r$ is the relative velocity between the liquid phase and gas phase, $a_L$ is the radius averaged throughout the ligament and $\rho_g$ is the ambient gas density. By taking $U_r \approx U_l$ and $a_L = 0.5<D_c>$, $We_g$ for the Jet A-1 spray from the CR orifice is found to be in the range of 3.5 to 7.9. Note that the actual value of $We_g$ is expected to be lower than the above estimated value as the local liquid phase velocity is always less than the liquid jet velocity at the orifice exit. This order of magnitude calculation supports the ligament/filament characteristics ($<D_c>$) recorded for the spray from the CR orifice. From Figure A2 (a) an increase in $We_l$ (or $U_l$) results in a decrease in the value of $<D_c>/D_{eq}$ for Jet A-1 jet from both circular and non-circular orifices. This behavior is attributed to the enhancement in atomization in the jet due to growth of surface instability with an increase in liquid inertia ($We_l$). Interestingly, the value of $<D_c>/D_{eq}$ for the fuel spray from the CR orifice is significantly lower compared to that obtained from the non-circular orifices. For a given orifice, the ligament/filament length, $L_c$ decreases with increase in $We_l$.

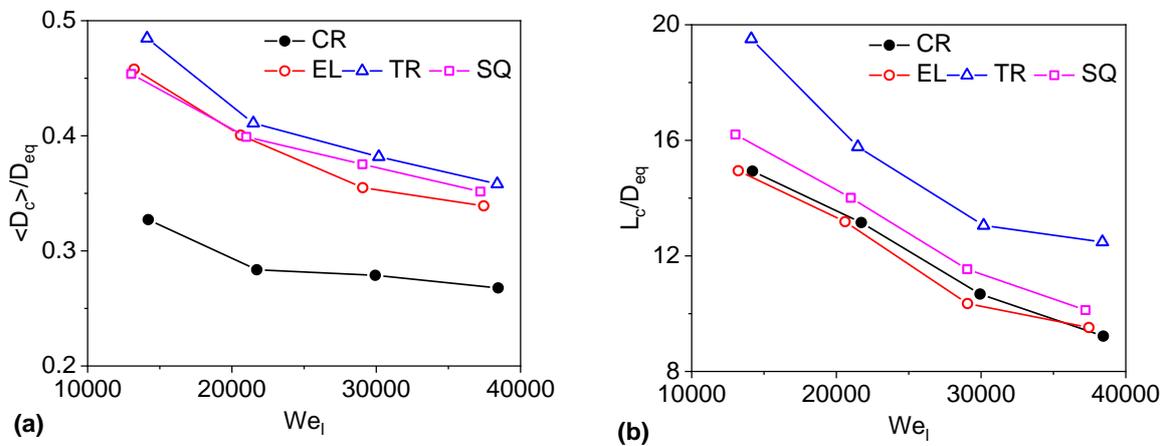

**Figure A2.** The variation of **(a)** $<D_c>/D_{eq}$, and **(b)** $L_c/D_{eq}$ with $We_l$ for the Jet A-1 fuel jet from the circular and non-circular orifices.

Among the non-circular orifices, one can observe that the orifice shape influences the value of $L_c/D_{eq}$ for a given flow condition as plotted in Figure A2 (b). For a given $We_l$, the liquid jet from the triangular orifice generates ligaments/filaments with higher $L_c/D_{eq}$ and that from the elliptical orifice generates ligaments/filaments with lower $L_c/D_{eq}$. This behavior of the ligament/filament characteristics is in contrast with the observed role of the orifice shape on the variation of $<D_c>/D_{eq}$ with $We_l$.

**Appendix 2: Ligament/Filament stability**

The experimental measurements $<D_c>$ and $L_c$ pave the way for the estimation of ligament/filament aspect ratio ($\Gamma$) given in Eq. (A1) by considering $0.5<D_c>$ and $L_c$ as the characteristic length scales of ligament/filament radius ($R_l$) and ligament/filament length ($L_l$), respectively. The calculated $\Gamma$ at different flow conditions can describe the collapse of the ligament/filament, at least qualitatively using Eqs. (A1) – (A6) below for smooth ligaments. The collapse of a ligament/filament is determined by its viscosity, size, and geometrical shape details (Driessen T. et al. 2013; Eggers and Villermaux 2008). The ligament/filament aspect ratio, $\Gamma$, expressed as

$$\Gamma = \frac{0.5 L_l}{R_l},$$
(A1)

where $L_l$ and $R_l$ are the length and radius of the ligament/filament, respectively, is a key parameter to determine the ligament/filament stability. According to Driessen et al. (Driessen T. et al. 2013), a smooth liquid ligament/filament collapses into droplets via Rayleigh-Plateau instability if $\Gamma$ is larger than a critical value, referred to as $\Gamma_c$. Through scaling analysis, Driessen et al.(Driessen T. et al. 2013) provided an analytical equation to determine $\Gamma_c$ and is expressed as

$$\frac{log(\epsilon)}{t_c \omega_{max}(Oh_l)} + (6\Gamma_c)^{\frac{1}{3}} - \Gamma_c = 0 \qquad (A2)$$

$$Oh_L = \frac{\mu}{\sqrt{\rho \sigma R_L}} \qquad (A3)$$

$$t_c \omega_{max}(Oh_L) = \sqrt{\frac{1}{2}(\kappa_{max}^2 - \kappa_{max}^4) + \frac{9}{4} Oh_L^2 \kappa_{max}^4} - \frac{3}{2} Oh_L \kappa_{max}^2 \quad \text{and} \qquad (A4)$$

$$\kappa_{max} = \left(2 + 3\sqrt{2} Oh_L\right)^{-\frac{1}{2}}, \qquad (A5)$$

where $Oh_L$ is the Ohnesorge number of the ligament/filament, and $\varepsilon$ is the relative initial distortion. The value of $\varepsilon$ is estimated from the initial amplitude of the perturbation, $\delta_L$ as,

$$\epsilon = \frac{\delta_L}{R_L}. \qquad (A6)$$

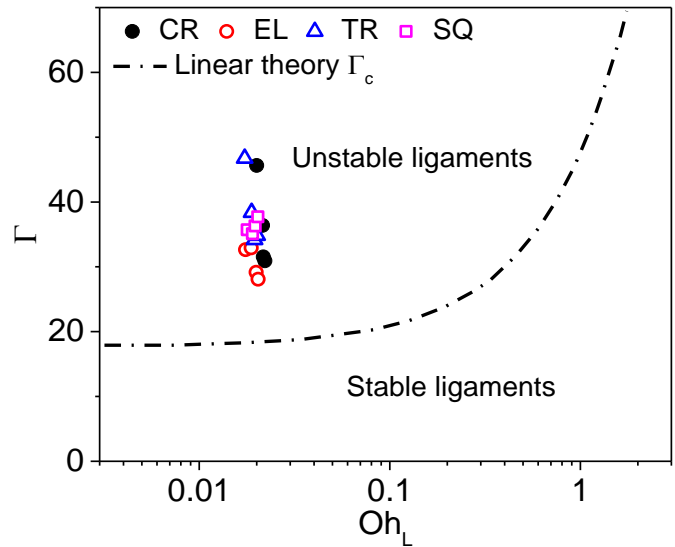

**Figure A3.** The variation of $\Gamma$ with $Oh_L$ for Jet A-1 jets from the circular and non-circular orifices. The variation of $\Gamma_c$ versus $Oh_L$ obtained from Eq. (A2) is shown as dash-dot line.

Note that the experimental determination of $\varepsilon$ is difficult and the analysis often involves the assuming a certain value for it.

Figure A3 shows the variation of $\Gamma$ with Ohnesorge number of the ligament ($Oh_L$) for the Jet A-1 jet from the circular and non-circular orifices. It also highlights the variation of $\Gamma_c$ (critical $\Gamma$) with $Oh_L$ obtained using Eqs. (A2) – (A6) as dash-dot line. Note that the theoretical analysis considers symmetric ligament/filament with negligible relative initial distortion ($\varepsilon$) whereas the ligaments/filaments recorded in the current study of Jet A-1 jets are asymmetric with significant distortion levels. This limitation restricts the current discussion on the stability of ligaments however our current comparison serve to provide a bound for the stability of these ligaments. It can be seen from Figure A2 that the ligament/filament formed from the circular and non-circular orifices are unstable, leading to the fragmentation of ligaments into droplets, as the measurements in the figure lie above the dash-dot line. This supports the complete atomization of the Jet A-1 jets at these flow conditions as highlighted in Figure 3.

Figure A2 suggests that, among the non-circular orifices, the triangular orifice produces the most unstable ligament/filament, a conclusion drawn based on the value of $\Gamma$ deviation from the $\Gamma_c$ line. This finding perhaps hints at faster atomization of the Jet A-1 jet from the triangular orifice compared to that from the elliptical and square orifices, which is discussed further in the next section of droplet size characteristics. In a similar manner, the elliptical orifice produces the least unstable ligament/filament, which hints at slower atomization of Jet A-1 jet from the elliptical orifice compared to the triangular and square orifices.

**Appendix 3: Validation with drop sizes generated from circular orifices in literature.**

Figure A4 shows the variation of maximum SMD at a given $We_l$, referred to as $SMD_{max}$, with $We_l$ for the Jet A-1 jet from the CR orifice. (Hiroyasu and Arai 1990) carried out systematic measurements of droplet size distribution in plain orifice atomizers using laser diffraction theory equipment and analysed the effects of various parameters, like the ambient gas pressure, liquid injection pressure, nozzle geometrical parameters and liquid properties, on the SMD. An empirical model is proposed in the study for the calculation of SMD in terms of atomizer geometrical parameters, flow conditions, liquid properties, and ambient gas properties. The empirical correlation of the SMD is expressed as (Hiroyasu and Arai 1990)

$$SMD = max(SMD_{LS}, SMD_{HS})$$
(A7)

$$SMD_{LS} = 4.12 D_o Re_l^{0.12} We_l^{-0.75} \left(\frac{\mu_l}{\mu_g}\right)^{0.54} \left(\frac{\rho_l}{\rho_g}\right)^{0.18}$$
(A8)

$$SMD_{HS} = 0.38 D_o Re_l^{0.25} We_l^{-0.32} \left(\frac{\mu_l}{\mu_g}\right)^{0.37} \left(\frac{\rho_l}{\rho_g}\right)^{-0.47}.$$
(A9)

The variation of SMD for the Jet A-1 jet from the CR orifice predicted using Eqs. A7 – A9 is shown in Figure A4 as solid line. The current experimental measurements of SMD with $We_l$ (symbols in Figure. A4) agree well with the predictions obtained from the empirical model by (Hiroyasu and Arai 1990).

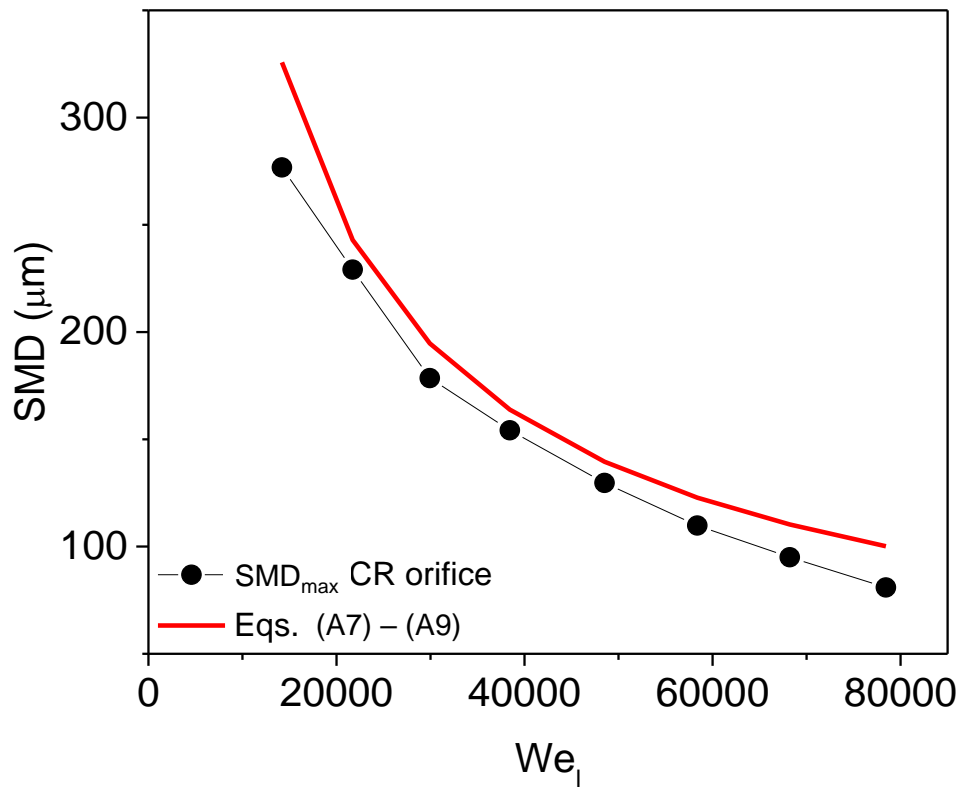

**Figure A4.** The comparison of current experimental measurements of SMD for the Jet A-1 jet from the CR orifice with the predicted value of SMD obtained from the empirical model by (Hiroyasu and Arai 1990) .

# References

Abramaoff, M. D., P. J. Magalhaes, and S. J. Ram. 2004. 'Image Processing Using ImageJ.' *Biophotonics International* 11:36–42.

Akinyemi, O. S., I. Qavi, C. E. Taylor, and L. Jiang. 2023. 'Effect of the Air-to-Liquid Mass Ratio on the Internal Flow and near-Field Spray Characteristics of a Two-Phase Swirl Burst Injector'. *Journal of Aerosol Science* 167:106092.

Amini, G., and A. Dolatabadi. 2011. 'Capillary Instability of Elliptic Liquid Jets'. *Physics of Fluids* 23:084109.

Amini, G., and A. Dolatabadi. 2012. 'Axis-Switching and Breakup of Low-Speed Elliptic Liquid Jets'. *International Journal of Multiphase Flow* 42:96–103.

Amini, G., Y. Lv, A. Dolatabadi, and M. Ihme. 2014. 'Instability of Elliptic Liquid Jets: Temporal Linear Stability Theory and Experimental Analysis'. *Physics of Fluids* 26:114105.

Baik S., Blanchard J.P., and Corradini M.L. 2003. 'Development of Micro-Diesel Injector Nozzles via MEMS Technology and Effects on Spray Characteristics'. *Atom. Sprays* 13:443-474.

Beale, J. C., and R. D. Reitz. 1999. 'Modeling Spray Atomization with the Kelvin-Helmholtz/Rayleigh-Taylor Hybrid Model'. *Atomization and Sprays* 9(6).

Bechtel, S. E. 1989. 'The Oscillation of Slender Elliptical Inviscid and Newtonian Jets: Effects of Surface Tension, Inertia, Viscosity and Gravity'. *Journal of Applied Mechanics* 56:968–74.

Bechtel S.E., Lin K.J., and Forest M.G. 1988. 'On the Behavior of Viscoelastic Free Jets with Elliptical Cross Section'. *Journal of Non Newtonian Fluid Mechanics* 27:87–126.

Bifeng, Y., Z. Ye, H. Jia, S. Yu, and W. Deng. 2022. 'Experimental Study on the Penetration of Diesel and Biodiesel Spray Liquid Emerging from an Equilateral Triangular Orifice under Evaporative Conditions'. *Journal of Thermal Science* 31:1–10.

Castrejon-Pita, J. R., S. D. Hoath, and I. M. Hutchings. 2012. 'Velocity Profiles in a Cylindrical Liquid Jet by Reconstructed Velocimetry. '. *Journal of Fluids Engineering* 134(1).

Chen, C., S. Yu, B. Yin, Q. Bi, and H. Jia. 2022. 'Investigation of Inner Flow and Near-Field Spray Patterns of the Non-Circular Diesel Injector'. *Sadhana* 47:1–8.

Chen S., and Ashgriz N. 2022. 'Droplet Size Distribution in Swirl Nozzles'. *International Journal of Multiphase Flow* 156:104219.

Chen, Siyu, and Nasser Ashgriz. 2022. 'Droplet Size Distribution in Swirl Nozzles'. *International Journal of Multiphase Flow* 156:104219.

Cho, M., S. Lee, M. Choi, and J. W. Lee. 2018. 'Novel Spray Tower for CO2 Capture Using Uniform Spray of Monosized Absorbent Droplets'. *Industrial & Engineering Chemistry Research* 57.(8):3065–75.

Driessen T., Jeurissen R., Wijshoff H., Toschi F., and Lohse D. 2013. 'Stability of Viscous Long Liquid Filaments.' *Physics of Fluids* 25:062109.

Dumouchel, C. 2008. 'On the Experimental Investigation of Primary Atomization of Liquid Streams'. *Experiments in Fluids* 45:371–442.

Eggers, J., and E. Villermaux. 2008. 'Physics of Liquid Jets'. *Reports on Progress in Physics* 71(3):036601.

Ehteram, M. A., H. B. Tabrizi, G. Ahmadi, M. Safari, and M. A. Mirsalim. 2013. 'Investigation of Fine Droplet Generation from Hot Engine Oil by Impinging Gas Jets onto Liquid Surface. '. *Journal of Aerosol Science* 65:49–57.

Farvardin, E., and A. Dolatabadi. 2013. 'Numerical Simulation of the Breakup of Elliptical Liquid Jet in Still Air'. *Journal of Fluids Engineering* 135:071302.

Gu, S., L. Wang, and D. L. Hung. 2017. 'Instability Evolution of the Viscous Elliptic Liquid Jet in the Rayleigh Regime'. *Physical Review E* 95:063112.

Hewitt, A. J. 1993. 'Droplet Size Spectra Produced by Air-Assisted Atomizers'. *Journal of Aerosol Science* 24(2):155–62.

Hiroyasu, H., and M. Arai. 1990. 'Structures of Fuel Sprays in Diesel Engines'. *SAE Transactions* 99:1050–61.

Jaberi, A., and M. Tadjfar. 2020. 'Comparative Study on Interfacial Oscillations of Rectangular and Elliptical Liquid Jets. '. *Proceedings of the Institution of Mechanical Engineers, Part G: Journal of Aerospace Engineering* 234:1272-1286.

Jia, H. ,., Y. ,. Jian, B. ,. Yin, J. Yang, and Z. ,. Liu. 2023. 'Experimental Study on the Combustion, Emissions and Fuel Consumption of Elliptical Nozzle Diesel Engine'. *Energy* 262:125449.

Jiang, Y., L. Hong, C. Chao, H. Lin, and Z. Daming. 2019. ' Hydraulic Performance and Jet Breakup Characteristics of the Impact Sprinkler with Circular and Non-Circular Nozzles'. *Applied Engineering in Agriculture* 35(6):911–24.

Jiayu, L., A. Leavey, Y. Wang, C. O'Neil, M. A. Wallace, C. A. D. Burnham, C. A. M. Boon, H. Babcock, and P. Biswas. 2018. 'Comparing the Performance of 3 Bioaerosol Samplers for Influenza Virus'. *Journal of Aerosol Science* 115:133–45.

Kasyap, T. V., Sivakumar D., and Raghunandan B.N. 2009. 'Flow and Breakup Characteristics of Elliptical Liquid Jets'. *International Journal of Multiphase Flow* 35:8–19.

Kasyap T.V., Sivakumar D., and Raghunandan B.N. 2008. 'Breakup of Liquid Jets Emanating from Elliptical Orifices at Low Flow Conditions'. *Atomization and Sprays* 18:1–24.

Kayahan, E., W. Min, T. Van Gerven, L. Braeken, S. Lambert, C. Politis, and M. E. Leblebici. 2022. 'Droplet Size Distribution, Atomization Mechanism and Dynamics of Dental Aerosols'. *Journal of Aerosol Science* 166:106049.

Kinnersley, R. P., and C. S. Parkin. 1995. 'Production of Droplets from Ligaments of Fluid in Increasingly Strong Normal Airflows'. *Journal of Aerosol Science* 8(26):1326.

Kulkarni, V., D. Sivakumar, C. Oommen, and T. J. Tharakan. 2010. 'Liquid Sheet Breakup in Gas-Centered Swirl Coaxial Atomizers'. *Journal of Fluids Engineering* 132(1).

Kulkarni, V., and P. E. Sojka. 2014. 'Bag Breakup of Low Viscosity Drops in the Presence of a Continuous Air Jet'. *Physics of Fluids* 26(7):072103.

Kumar, S. S., C. Li, C. E. Christen, C. J. Hogan Jr, S. A. Fredericks, and J. Hong. 2019. 'Automated Droplet Size Distribution Measurements Using Digital Inline Holography. '. *Journal of Aerosol Science* 137:105442.

Lefebvre, Arthur H., and Vincent G. McDonnel. 2017. *Atomization and Sprays*. CRC Press.

Longest, P. W., and T. H. Landon. 2012. 'In Silico Models of Aerosol Delivery to the Respiratory Tract—Development and Applications'. *Advanced Drug Delivery Reviews* 64(4):296–311.

Makhnenko, I., E. R. Alonzi, S. A. Fredericks, C. M. Colby, and C. S. Dutcher. 2021. 'A Review of Liquid Sheet Breakup: Perspectives from Agricultural Sprays. '. *Journal of Aerosol Science* 157:105805.

Marmottant P., and Villermaux E. 2004. 'On Spray Formation'. *Journal of Fluid Mechanics* 498:73-111.

Middleman, S. 1964. 'Profile Relaxation in Newtonian Jets'. *Industrial & Engineering Chemistry Fundamentals* 3(2):118–22.

Morad, M. R., M. Nasiri, and G. Amini. 2020. 'Numerical Modeling of Instability and Breakup of Elliptical Liquid Jets'. *AIAA Journal* 58:2442–49.

Muthukumaran, C. K., and A. Vaidyanathan. 2014. 'Experimental Study of Elliptical Jet from Sub to Supercritical Conditions'. *Physics of Fluids* 26:044104.

Nurick, W. H. 1976. 'Orifice Cavitation and Its Effect on Spray Mixing. '. *Journal of Fluids Engineering* 98:681–87.

Rayleigh, L. 1879. 'On the Capillary Phenomena of Jets'. *Proceedings of Royal Society of London* 29:71–97.

Raza, M. S., and K. A. Sallam. 2023. 'Primary Breakup of Liquid Fan Sheet in Crossflow. '. *Journal of Aerosol Science* 106135.

Sallam K.A., Dai Z., and Faeth G.M. 2002. 'Liquid Breakup at the Surface of Turbulent Round Liquid Jets in Still Gases'. *International Journal of Multiphase Flow* 28:427–49.

Sharma P, and Fang T. 2014. 'Breakup of Liquid Jets from Non-Circular Orifices'. *Exp. Fluids* 55:1-17.

Sharma P, and Fang T. 2015. 'Spray and Atomization of a Common Rail Fuel Injector with Non-Circular Orifices'. *Fuel* 153:416–30.

Shinjo, J., and S. Umemura. 2010. 'Simulation of Liquid Jet Primary Breakup: Dynamics of Ligament and Droplet Formation.' *International Journal of Multiphase Flow* 36:513–32.

Sivakumar, D., and V. Kulkarni. 2011. 'Regimes of Spray Formation in Gas-Centered Swirl Coaxial Atomizers'. *Experiments in Fluids* 51(3):587–96.

Sivakumar, D., S. K. Vankeswaram, R, Sakthikumar, and B. N. Raghunandan. 2015. 'Analysis on the Atomization Characteristics of Aviation Biofuel Discharging from Simplex Swirl Atomizer'. *International Journal of Multiphase Flow* 72:88–96.

Sivakumar D., Vankeswaram. S.K., R. Sakthikumar, Raghunandan. B.N., J. T. C. Hu, and A. K. Sinha. 2016. 'An Experimental Study on Jatropha-Derived Alternative Aviation Fuel Sprays from Simplex Swirl Atomizer'. *Fuel* 179:36–44.

Sou A., Hosokawa S., and Tomiyama A. 2007. 'Effects of Cavitation in a Nozzle on Liquid Jet Atomization. '. *International Journal of Heat and Mass Transfer* 50:3575–82.

Van Strien, J., P. Petersen, P. Lappas, L. Yeo, A. Rezk, S. Vahaji, and K. Inthavong. 2022. 'Spray Characteristics from Nasal Spray Atomization'. *Journal of Aerosol Science* 165:106009.

Triballier, C., C. Dumouchel, and J. Cousin. 2003. 'A Technical Study on the Spraytec Performances: Influence of Multiple Light Scattering and Multi-Modal Drop-Size Distribution Measurements'. *Experiments in Fluids* 35:347–56.

Vankeswaram, S. K. ,., and S. Deivandren. 2022. 'Size and Velocity Characteristics of Spray Droplets in Near-Region of Liquid Film Breakup in a Swirl Atomizer'. *Experimental Thermal and Fluid Science* 130:110505.

Vankeswaram, S. K., R. Sakthikumar, S. Deivandren, and J. D. Hu. 2023. 'Evaluation of Spray Characteristics of Aviation Biofuels and Jet A-1 from a Hybrid Airblast Atomizer'. *Experimental Thermal and Fluid Science* 142:110820.

Varga, C. M., J. C. Lahseras, and E. J. Hopfinger. 2003. 'Initial Breakup of a Small-Diameter Liquid Jet by a High-Speed Gas Stream'. *Journal of Fluid Mechanics* 497:405–34.

Villermaux, E. 2007. 'Fragmentation'. *Annual Review of Fluid Mechanics* 39:419–46.

Villermaux, E., P. Marmottant, and J. Duplat. 2004. 'Ligament-Mediated Spray Formation'. *Physical Review Letters* 92:074501(1-4).

Wang F, and Fang T. 2015. 'Liquid Jet Breakup for Non-Circular Orifices under Low Pressures.' *International Journal of Multiphase Flow* 72:248-262.

Xu, Y. C., Y. B. Li, Y. Z. Liu, Y. Luo, G. W. Chu, L. L. Zhang, and J. F. Chen. 2019. 'Liquid Jet Impaction on the Single-layer Stainless Steel Wire Mesh in a Rotating Packed Bed Reactor'. *AIChE Journal* 65(6):16597.

Yin, B., B. Xu, H. Jia, and S. Yu. 2020. 'The Effect of Elliptical Diesel Nozzles on Spray Liquid-Phase Penetration under Vaporative Conditions'. *Energies* 13(2234):1–14.

Yu S., Yin B., Deng W., Jia H., Ye Z., Xu B, and Xu H. 2018. 'Experimental Study on the Spray Characteristics Discharging from Elliptical Diesel Nozzle at Typical Diesel Engine Conditions'. *Fuel* 221:28-34.

Yu S, Yin B, Deng W, Jia H, Ye Z, Xu B, and H. Xu. 2018. 'Internal Flow and Spray Characteristics for Elliptical Orifice with Large Aspect Ratio under Typical Diesel Engine Operation Conditions.' *Fuel* 228:62-73.

Yu S., Yin B., Deng W., Jia H., Ye Z., Xu B., and Xu H. 2019. ' An Experimental Comparison of the Elliptical and Circular Nozzles Spray and Mixing Characteristics under Different Injection Pressures. '. *Fuel* 236:1474–82.

Yu, S., B. Yin, W. Deng, H. Jia, Z. Ye, B. Xu, and H. Xu. 2019. 'Experimental Study on the Spray and Mixing Characteristics for Equilateral Triangular and Circular Nozzles with Diesel and Biodiesel under High Injection Pressures'. *Fuel* 239:97–107.

Yu S., Yin B., Jia H., Chen C., and Xu B. 2021. 'Investigation of Inner Cavitation and Nozzle Exit Flow Patterns for Elliptical Orifice GDI Injectors with Various Aspect Ratios'. *International Communications in Heat and Mass Transfer* 129:105682.